\documentclass[11pt]{article}

\usepackage{amsmath}
\usepackage{amsthm}
\usepackage{amssymb}
\usepackage{amsfonts}
\usepackage{graphicx}

\usepackage{hyperref}
\usepackage{float}
\usepackage{color}
\def\f12{\frac 1 2}

\def\hh{\mathcal{H}^{+}}

\def\f12{\frac 1 2}

\newtheorem{theorem}{Theorem}[section]

\begin{document}
\title{Nonlinear instability of scalar fields on extremal black holes}


\author{Stefanos Aretakis\thanks{Princeton University, Department of Mathematics, Fine Hall, Washington Road, Princeton, NJ 08544, USA.}\thanks{  Institute for Advanced Study, Einstein Drive, Princeton, NJ 08540, USA.}}

\maketitle

\begin{abstract}
We prove a new type of finite time blow-up for a class of semilinear wave equations on extremal black holes. The initial data can be taken to be arbitrarily close to the trivial data.  The first singularity occurs along the (degenerate) future event horizon. No analogue of this instability occurs for subextremal black holes or the Minkowski spacetime. 

\end{abstract}

\section{Introduction}
\label{sec:Introduction}

The dispersion of the linear wave equation 
\begin{equation}
\Box_{g}\psi=0
\label{we}
\end{equation} on black hole spacetimes $(\mathcal{M},g)$ provides a good indication of the stability properties of the background in the context of the Cauchy problem of the Einstein equations. Definitive quantitative decay results for solutions to the wave equation \eqref{we} in the exterior of general sub-extremal Kerr backgrounds $(|a|<M)$ were proven in \cite{tria, yakov2} (see also \cite{dr7,enadio,tataru2, blukerr} for small slowly rotating Kerr spacetimes). These works, however, did not include the extremal case. 

The general study of the wave equation \eqref{we} on extremal black holes was initiated by the author in a series of papers  \cite{aretakis1,aretakis2,aretakis3,aretakis4,aretakis2012} where it was shown that solutions to \eqref{we} exhibit both stability and instability properties. In particular, it was shown that $|\psi|$ decays in the future; however, there are translation-invariant derivatives $Y\psi$ that generically do not decay along the future event horizon $\hh$. Moreover, the higher order derivatives $Y^{k}\psi$, $k\geq 2$ were shown to asymptotically blow up along $\hh$. 

 \begin{figure}[H]
	\centering
		\includegraphics[scale=0.127]{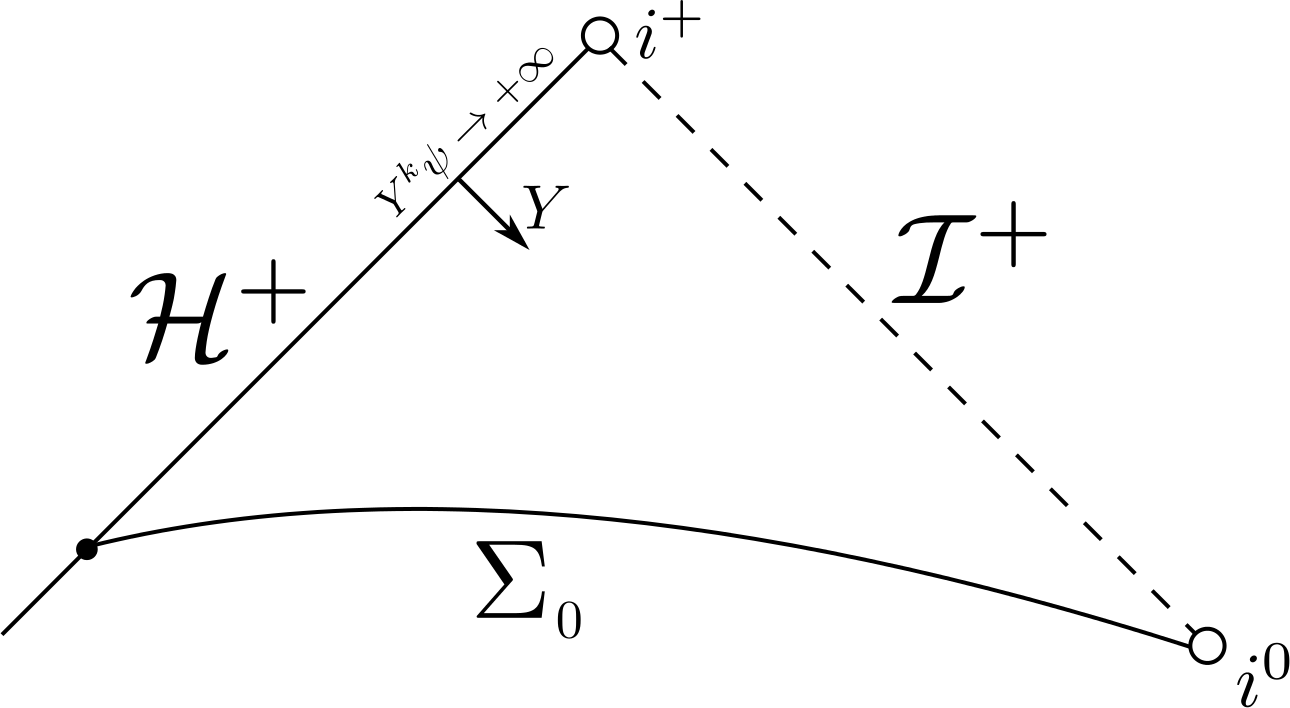}
	\label{fig:ekerr2cd1}
	\caption{The derivatives $Y^{k}\psi$ asymptotically blow up along the event horizon}
\end{figure}

The source of these instabilities is a hierarchy of conservation laws for the scalar field along the event horizon first established in the aforementioned series of papers. These conservation laws  have been extended to more general linear fields such as Maxwell's equations and linearized gravity on extremal Kerr by Lucietti and Reall \cite{hj2012}. Murata \cite{murata2012} has also provided similar generalizations for extremal vacuum black holes in arbitrary dimensions.  Bizon and Friedrich \cite{bizon2012} have remarked that the conserved quantities  on exactly extremal Reissner--Nordstr\"{o}m correspond to the Newman--Penrose constants at null infinity under a conformal transformation of the background that  exchanges the (future) event horizon with (future) null infinity.  The relation between the conserved quantities of the present paper and the Newman--Penrose constants was also independently observed by Lucietti, Murata, Reall and Tanahashi \cite{hm2012}. The same authors studied analytically and numerically the late time behavior of massive and massless scalars on extremal Reissner--Nordstr\"{o}m. An important conclusion of their numerical analysis is that scalar instabilities are present even if the scalar perturbation is initially supported away from the horizon (in which case all the conserved quantities are zero).  The author  rigorously showed in \cite{aretakis2012} that  perturbations which are initially supported away from the horizon  indeed (generically) develop instabilities in the future confirming the numerical analysis of \cite{hm2012}.

The study, however, of purely linear fields is not satisfactory in the context of the Cauchy problem to the Einstein equations. The simplest non-linear problem one can consider is that of a semi-linear wave equation
\begin{equation}
\Box_{g}\psi=N(\psi,\partial\psi),
\label{nwe}
\end{equation}
where $N(\psi,\partial\psi)$ is a non-linear expression of $\psi$ and $\partial\psi$. The study of such equations is of course more complicated even for the flat Minkowski spacetime $(\mathbb{R}^{3+1},m)$. It is well known that if the non-linearity satisfies
\begin{equation}
N(\psi,\partial\psi)=O(|\psi|^{n})+O(|\partial\psi|^{n}), \ \ n\geq 3, 
\label{eq:}
\end{equation}
then small data lead to global solutions in time. Note that the power here must satisfy $n\geq 3$. This is because the decay for $\psi,\partial\psi$ is not very strong and hence a sufficiently high power in the non-linearity is needed in order to stabilize the evolution. On the other hand, the quadratic case where $n=2$ is borderline. Indeed, John \cite{john} showed that any nontrivial $C^{3}$ solution of the wave equation 
\[\Box_{m}\psi=(\partial_{t}\psi)^{2}\]
blows up in finite time. Moreover, Klainerman \cite{sergiunull} showed that if the nonlinearity $N$ satisfies the so-called null condition then global existence  is guaranteed for small data. The nonlinearity $N=N(\psi,\partial\psi)$ satisfies the null condition with respect to the Minkowski metric if 
\begin{equation}
\begin{split}
N(\psi,\partial\psi)=A^{\alpha\beta}\partial_{\alpha}\psi\partial_{\beta}\psi,
\end{split}
\label{nullc}
\end{equation}
where  $A^{\alpha\beta}$ are constants and satisfy $A^{\alpha\beta}\xi_{\alpha}\xi_{\beta}=0$ whenever $\xi$ is a null vector. 

Recently, Yang \cite{shiwu} has in fact extended the above result to more general perturbations of the Minkowski spacetime using a robust method introduced by Dafermos and Rodnianski \cite{newmethod}. His result can in fact be modified to work for Schwarzschild and more general sub-extremal Kerr backgrounds. Luk \cite{luknullcondition} has shown a similar result for slowly rotating Kerr backgrounds. 

The aim of this paper is to show that there exist nonlinearities $N=N(\psi,\partial\psi)$ on extremal Kerr such that 
\begin{equation*}
\begin{split}
&N(0,0)=0, \ \ \ dN(0,0)=0,\\
N(\psi&,\partial\psi)=O(|\psi|^{n})+O(|\partial\psi|^{n}),\ \  n\geq 2
\end{split}
\end{equation*}
 and
for which $C^{1}$ solutions to \eqref{nwe} with arbitrarily small initial data blow up in finite time. It is important to emphasize here that $n\in\mathbb{N}$ can be chosen arbitrarily large (and hence this result is in stark contrast with the subextremal case).  As we shall see, this result holds for more general extremal black holes. This shows that extremal black holes exhibit a genuine non-linear scalar instability.

\section{A genuine non-linear instability}
\label{sec:ANonLinearInstability}

Let $(\mathcal{M},g)$ denote the exterior region of an extremal Kerr black hole. Let $\Sigma_{0}$ be a spacelike hypersurface which  that the future event horizon $\hh$ and terminates at spacelike infinity $i^{0}$.
We consider the following equation:
\begin{equation}
\Box_{g}\psi=N(\psi,\partial\psi)
\label{eq:1}
\end{equation}
where 
\begin{equation}
N(\psi,\partial\psi)=\psi^{2n}+(Y\psi)^{2n}+(T\psi)^{2n},
\label{ton}
\end{equation}
where $T=\partial_{v}$, $Y=\partial_{r}$ and $n\in\mathbb{N}$ with $n\geq 1$. Here $\partial_{v},\partial_{r}$ correspond to partial derivatives with respect to the ingoing Eddington--Finkelstein coordinates $(v,r,\theta,\phi^{*})$.

 We also prescribe smooth compactly supported initial data 
\[\psi_{0}=\left.\psi\right|_{\Sigma_{0}}, \ \ \ \psi_{1}=\left.n\psi\right|_{\Sigma_{0}}, \]
where $n$ denotes the unit normal to $\Sigma_{0}$. 

The main result of the present paper is the following:

\begin{theorem} For all natural numbers $n\geq 1$ there are solutions of \eqref{eq:1} with arbitrarily small initial data $(\psi_{0},\psi_{1})$ which in finite time fail to be $C^{1}$. In other words, there are solutions of \eqref{eq:1} with arbitrarily small initial data such that $N(\psi, \partial\psi)\rightarrow+\infty$ in finite time along the future event horizon $\hh$.
\label{theo}
\end{theorem}

 \begin{figure}[H]
	\centering
		\includegraphics[scale=0.127]{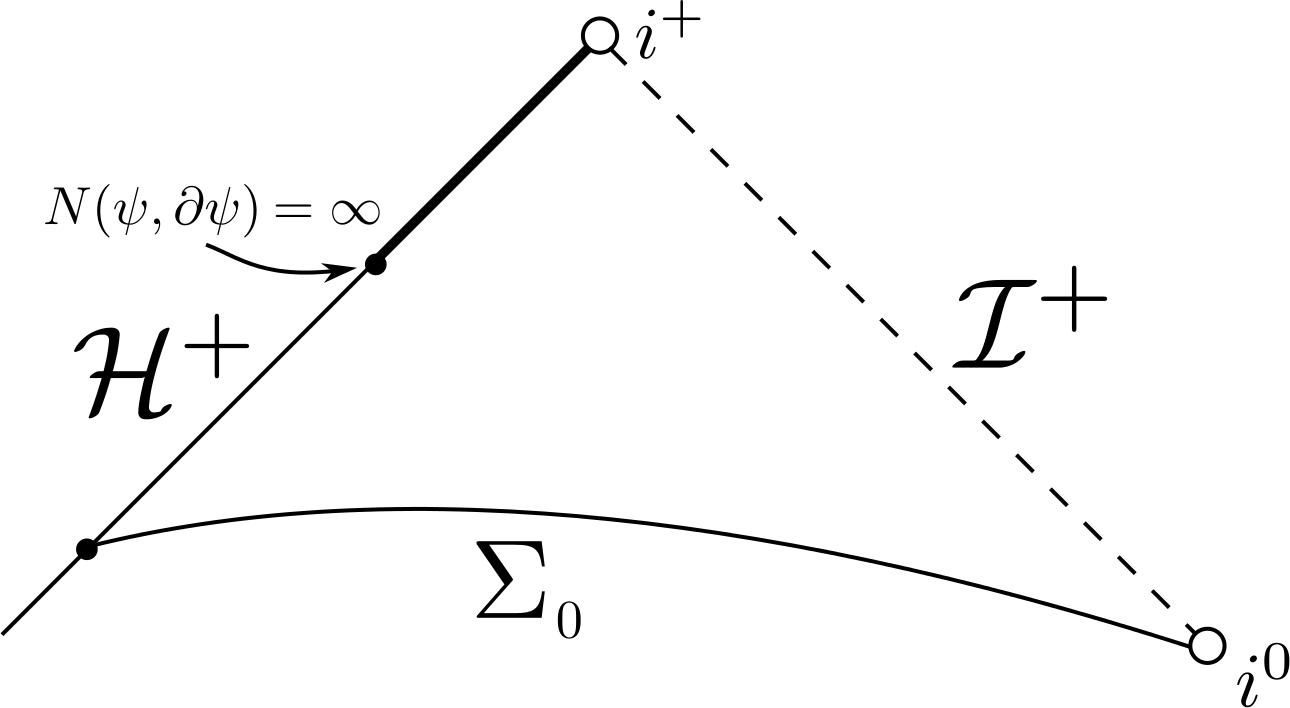}
	\label{fig:ekerr2cd12}
	\caption{The term $N(\psi,\partial\psi)$ blows up at finite time along the event horizon}
\end{figure}

\begin{proof}
Let $\Sigma_{\tau}=\phi^{T}_{\tau}(\Sigma_{0})$, where $\phi^{T}_{\tau}$ denotes the flow of $T$. Let also $S_{\tau}=\Sigma_{\tau}\cap\hh$. Note that $S_{\tau}$ is diffeomorphic to $\mathbb{S}^{2}$. Consider initial data $(\psi_{0},\psi_{1})$ such that 
\[H_{0}[\psi]=\int_{S_{0}}M\sin^{2}\theta (T\psi)+4M(Y\psi)+2\psi\  =\epsilon,\]
where $\epsilon>0$ can be taken to be arbitrarily small. 

In \cite{aretakis4} it was established that for linear scalar perturbations there exists a conserved quantity $H_{0}[\psi]$ along $\hh$. In our case, the existence of this quantity implies that if $\psi$ satisfies \eqref{eq:1} then 
\begin{equation}
T(H_{\tau})=\int_{S_{\tau}}N,
\label{eq:2}
\end{equation}
where
\[H_{\tau}[\psi]=\int_{S_{\tau}}M\sin^{2}\theta (T\psi)+4M(Y\psi)+2\psi\] and  $N=N(\psi,\partial\psi)$ is given by \eqref{ton}. Note now the following:
\begin{equation}
\begin{split}
\int_{S_{\tau}}N&=\int_{S_{\tau}}\psi^{2n}+(Y\psi)^{2n}+(T\psi)^{2n}\geq\int_{S_{\tau}}\psi^{2n}+(Y\psi)^{2n}+(\sin\theta)^{4n}\cdot(T\psi)^{2n} \\ &\geq \tilde{C}\int_{S_{\tau}}\Big(\psi+Y\psi+(\sin\theta)^{2}\cdot(T\psi)\Big)^{2n}\geq
C\,\Bigg(\int_{S_{\tau}}\psi+Y\psi+(\sin\theta)^{2}\cdot(T\psi)\Bigg)^{2n}\\&\geq C\, \Big(H_{\tau}\Big)^{2n}
\end{split}
\label{mak}
\end{equation}
where we have used Chebyshev's inequality 
\[a^{2n}+b^{2n}+c^{2n}\geq \tilde{C}\big(a+b+c\big)^{2n}, \ \ a,b,c\in\mathbb{R},\]
and H\"{o}lder's inequality
\[\Bigg(\int_{S_{\tau}}|f|^{2n}\Bigg)^{\frac{1}{2n}}\cdot\Bigg(\int_{S_{\tau}}1\Bigg)^{\frac{2n-1}{2n}}\geq \int_{S_{\tau}}|f|,\] which implies
\[ \int_{S_{\tau}}|f|^{2n}\geq C\, \Bigg(\int_{S_{\tau}}|f|  \Bigg)^{2n} \geq C\, \Bigg(\int_{S_{\tau}}f  \Bigg)^{2n}.  \] 
In view of \eqref{eq:2} and \eqref{mak} we obtain
\begin{equation}
T(H_{\tau})\geq C\, (H_{\tau})^{2n}\geq 0\, :\text{ on }\hh,
\label{onh}
\end{equation}
where $C$ is a constant that depends only on $n$ and the geometry of $S_{0}$. 
Since $\tau$ parametrizes the flow of $T$, we should think of $T$ as ``$\,\partial_{\tau}$''. 
Since we assume that initially $H_{0}=\epsilon>0$ and since, in view of \eqref{onh}, $H_{\tau}$ is non-decreasing, we obtain that 
\[H_{\tau}>0\, :\text{ for all }\tau\geq 0.\]
Hence we can rewrite \eqref{onh} as follows
\[T\left(-\frac{1}{(H_{\tau})^{2n-1}}\right)\geq C.\]
Therefore,
\[ \frac{1}{(H_{0})^{2n-1}}-\frac{1}{(H_{\tau})^{2n-1}}\geq C\cdot\tau\ \ \Rightarrow \ \ \frac{1}{(H_{\tau})^{2n-1}}\leq -C\cdot\tau+\frac{1}{\epsilon^{2n-1}}.\] 
Therefore, as $\tau\rightarrow \frac{1}{C\cdot \epsilon^{2n-1}}$ we necessarily have $H_{\tau}\rightarrow+\infty$. This immediately shows that $\psi$ cannot be extended as a $C^{1}$ function in the region $\tau\in \left[0,\frac{1}{C\cdot \epsilon^{2n-1}}\right]$.

\end{proof}

We remark that a modification of our arguments works for extremal Reissner--Nordstr\"{o}m and in fact, in view of the conservation laws established in \cite{aretakis4,hj2012}, it works for more general extremal black holes. Further study of the blow-up will be presented elsewhere.

\section{Acknowledgements}
\label{sec:Acknowlegments}
I would like to thank Mihalis Dafermos and Shiwu Yang for very helpful discussions.

\bibliographystyle{acm}
\bibliography{../../../bibliography}
\end{document}